\documentclass[12pt,preprint]{aastex}

\slugcomment{Accepted for publication in The Astrophysical Journal}

\shorttitle{Diameter of $\lambda$~Boo}

\shortauthors{Ciardi et al.}

\begin{document}

\title{
The Angular Diameter of $\lambda$ Bo\"{o}tis}

\author{David R. Ciardi, Gerard T. van Belle, Andrew F. Boden}
\affil{Michelson Science Center/Caltech\\
 770 South Wilson Avenue, M/S 100-22
Pasadena, CA 91125}
\email{ciardi@ipac.caltech.edu}

\author{T. ten Brummelaar, H. A. McAlister, W.G. Bagnuolo, Jr.\\
P. J. Goldfinger, J. Sturmann, L. Sturmann, N. Turner, } \affil{Center for
High Angular Resolution Astronomy, Department of Physics and Astronomy,
Georgia State University, Atlanta, GA 30302-4106}

\author{D. H. Berger}

\affil{University of Michigan, Dept.~of Astronomy,\\ Ann Arbor, MI 48109-1042}

\author{R. R. Thompson}
\affil{Michelson Science Center/Caltech\\
 770 South Wilson Avenue, M/S 100-22
Pasadena, CA 91125}

\and

\author{S. T. Ridgway}
\affil{National Optical Astronomy Observatories\\
P.O. Box 26732, Tucson, AZ 85726-6732}

\begin{abstract}

Using the CHARA Array and the Palomar Testbed Interferometer, the chemically
peculiar star $\lambda$ Bo\"{o}tis has been spatially resolved. We have
measured the limb darkened angular diameter to be $\theta_{LD} =
0.533\pm0.029$ mas, corresponding to a linear radius of $R_{\star} =
1.70\pm0.10$~$R_\odot$. The measured angular diameter yields an effective
temperature for $\lambda$~Boo of $T_{eff} = 8887 \pm 242$~K.  Based upon
literature surface gravity estimates spanning $\log{(g)} = 4.0-4.2$ $[\rm{cm\
s}^{-\rm{2}}]$, we have derived a stellar mass range of $M_{\star} = 1.1 -
1.7$ $M_\odot$.  For a given surface gravity, the linear radius uncertainty
contributes approximately $\sigma(M_\star) = 0.1-0.2\ M_\odot$ to the total
mass uncertainty. The uncertainty in the mass (i.e., the range of derived
masses) is primarily a result of the uncertainty in the surface gravity.  The
upper bound of our derived mass range ($\log(g)=4.2,\ M_\star = 1.7\pm0.2\
M_\odot$) is consistent with $100-300$ MYr solar-metallicity evolutionary
models.  The mid-range of our derived masses ($\log(g)=4.1,\ M_\star =
1.3\pm0.2\ M_\odot$) is consistent with $2-3$ GYr metal-poor evolutionary
models.  A more definitive surface gravity determination is required to
determine a more precise mass for $\lambda$~Boo.

\end{abstract}


\section{Introduction \label{intro-sec}}

$\lambda$~Bo\"otis stars are a chemically peculiar class of late-B to mid-F
stars \citep{mkk43}.  The stars are depleted of heavy elements like Mg and Fe
($[{\rm M}/{\rm H}] = -2.0$), but exhibit solar abundances for light elements
such as C, N, O, and S \citep[e.g.,][]{hs83,gray88,vl90}.   Approximately 2\%
of the known A-stars in the field have been classified as $\lambda$ Boo-type
stars \citep{gc02}.  On an HR diagram the $\lambda$ Boo stars appear to lie
between the zero-age and terminal age main sequences clouding the nature and
evolutionary status of these stars \citep{pg97,gc02}. \citet{solano01}
provides an introduction into the competing theories for the nature of the
$\lambda$ Boo stars, briefly summarized here.

The first hypothesis is that $\lambda$~Boo stars are young main sequence stars
which are still surrounded by a shell or disk of gas and dust \citep{vl90}.
The heavy, refractory elements are locked within the surrounding dust grains.
The volatile elements remain in the gas and accrete onto the star, while the
dust grains are blown away by the stellar radiation pressure taking the heavy
elements with them, requiring the presence of circumstellar dust.  All four of
the known $\lambda$ Boo stars within 40 pc have detected infrared excesses
\citep{gc02,jura05,rieke05,chen06} indicative of circumstellar dust.  A
continual accretion of the light gases at a rate of $\approx10^{-13}\ M_\odot\
yr^{-1}$ \citep{charbonneau93} is needed. Once the accretion stops, the
observed metal deficiencies fade within a million years. It is unclear if the
surrounding disks contain enough light element gases to sustain the needed
accretion rate over the main sequence lifetime of the star.

In this model, $\lambda$ Boo stars would be relatively young (tens to
hundreds of MYr) with solar-metallicity but with metal-deficient
photospheres. \citet{gc98} obtained spectra of 60 Herbig Ae and pre-main
sequence A-stars, which, in this scenario, would be expected to contain a
higher fraction of $\lambda$~Boo stars than the general field, but found only
one $\lambda$ Boo star and one marginal $\lambda$ Boo star, a rate comparable
to the field star rate.

A variation of this hypothesis places the $\lambda$~Boo stars at the end of
their main sequence lifetimes, and the shell is the result of mass loss.
After $10^9$ years of mass loss, diffusion in the atmosphere produces
underabundances of the heavier elements \citep{mc86}. However, it is not
clear if this mechanism can produce the strong underabundances of heavy
elements that is observed in the $\lambda$ Boo stars. In this hypothesis,
$\lambda$ Boo stars would be relatively old (a few GYr). At these ages, the
$\lambda$ Boo stars may be intrinsically more metal-poor than comparable
A-stars which are younger.

Interestingly, \citet{paunzen02} concluded that the field $\lambda$ Boo stars
are located relatively homogeneously throughout their main sequence
evolution.  Based upon comparison to solar-metallicity isochrones, they find
a uniform distribution of ages for $\lambda$ Boo stars between 10 Myr to 500
Myr.  This is followed by a rise in the number of $\lambda$ Boo stars at an
age of 0.6-1 Gyr, at which point the fraction of $\lambda$ Boo stars relative
to normal A-stars is higher than at younger ages.

An alternative hypothesis is that $\lambda$~Boo stars are binary stars with
both stars being of similar spectral type.  The composite spectrum produces
an apparent under-abundance of heavy elements \citep[e.g.,][]{fb99,gfl03}.  A
complementary proposal is that $\lambda$~Boo stars are actually contact
binary stars \citep{andrievsky97}.  The composite colors of the star would
look normal, but the spectral abundances would appear ``metal-poor''
\citep{fb05}.

Nearly all of the work on $\lambda$~Boo stars has involved detailed color
and/or spectral analysis of the stars to determine effective temperatures,
surface gravities, and elemental abundances. Determinations of basic stellar
parameters, such as the stellar radii and masses, have been made indirectly
from photometric fitting and comparison to evolutionary models. Optical
interferometry, which is capable of resolving the stellar disk can add
crucial and independent information to the debate on $\lambda$~Boo stars.

We have made the first direct measurements of the angular diameter of the
prototype for the class, $\lambda$~Bo\"otis \citep[HD 125162, A3 V
kB9.5mB9.5;][]{gray03}, using the Georgia State University's (GSU) Center for
High Angular Resolution Astronomy (CHARA) Array and the Palomar Testbed
Interfometer (PTI).  The CHARA Array with its long baselines ($200 - 300$ m)
is uniquely suited for observations of absolute diameters of main sequence
stars, thereby, providing a unique perspective on the evolutionary status of
$\lambda$~Boo.

\section{Observations and Data Reduction \label{obs-sec}}

$\lambda$~Boo was observed, in conjunction with two calibration stars, with
the CHARA Array at 2.2~$\mu$m on 4 nights between 2004 Jun 17 and 2004 Jun 29,
utilizing the W1-E1 and E1-S1 baselines.  It was then observed two years later
on 2006 Jun 29 and 2006 Jun 30 with the E1-S1 baseline at 1.67 \micron.
$\lambda$~Boo, along with the calibration stars HD 125349 and HD 129002, was
observed multiple times during each of these nights, and each observation, or
scan, was approximately 200 s long. Observations of both calibrators bracketed
each observation of $\lambda$~Boo.

For each scan, we computed a mean $V^2$-value from the scan data, and the
error in the $V^2$ estimate from the rms internal scatter \citep{ten05}.
$\lambda$~Boo was always observed in combination with its calibration sources
HD 125349 and HD 129002. The calibrators (see Table \ref{calib-tab}) are
expected to be unresolved by the interferometer with estimated angular sizes
of $0.198\pm0.012$ mas and $0.286\pm0.018$ mas, respectively. These angular
size estimates were based upon fitting template spectral energy distributions
(SED) of the proper spectral type from \citet{pic98} to available broadband
photometry available from IRSA\footnote{NASA's Infrared Science Archive} and
SIMBAD. These objects were additionally selected to be slow apparent rotators,
with $v \sin i <$ 30 km s$^{-1}$ to ensure the stars are circularly symmetric
\citep{ues82,henry00}.

The calibration of the $\lambda$~Boo $V^2$ data is performed by estimating
the interferometer system visibility ($V_{sys}^2$) using the calibration
source with model angular diameters and then normalizing the raw
$\lambda$~Boo visibility by $V_{sys}^2$ to estimate the $V^2$ measured by an
ideal interferometer at that epoch \citep{moz91,boden98}. Uncertainties in
the system visibility and the calibrated target visibility are inferred from
internal scatter among the data in a scan and standard error-propagation
calculations. More detail on the CHARA target and calibrator selection, data
reduction, and technical aspects for the CHARA Array is available in the
literature \citep{mca05,ten05,vanbelle06}.

In addition to the CHARA Array data, observations of $\lambda$~Boo were
obtained from the Palomar Testbed Interferometer \citep[PTI;][]{colavita99}
archive\footnote{The archive is available at the Michelson Science Center
(http://msc.caltech.edu).}. $\lambda$~Boo was observed with PTI in 2000,
2003, and 2004 with the N-S, N-W, and S-W baselines (85-100~m) at both K and
H bands. The PTI observations utilized the same calibrators as the CHARA
observations.

Keeping the CHARA and PTI data separate, the data were grouped by baseline.
The CHARA data were binned such that the bin widths were $<2\%$ of the central
baseline length. The PTI data were binned by baseline configuration (e.g.,
N-S) and by wavelength (K-band vs. H-band). For each bin, the mean baseline
lengths, position angles, and effective wavelengths were calculated, weighted
by the quality of the $V^2$ measurements. An error-weighted mean $V^2$ was
calculated for each bin. The resulting data are presented in Table
\ref{meanv2-tab}, and the resulting visibility plot is shown in Figure
\ref{v2-fig}.

\section{Discussion \label{disc-sec}}
The primary result of this paper is the measurement of the apparent angular
diameter for $\lambda$~Boo. In the following sections, we discuss the angular
diameter determination and the associated linear radius of $\lambda$~Boo.  We
then relate these measurements to the effective temperature and mass,
comparing $\lambda$~Boo to other A-stars.

\subsection{Angular Diameter \label{angd-subsec}}

We have modelled the observed mean visibilities as listed in Table
\ref{meanv2-tab} with a uniform disk of angular size $\Theta_{UD} $ of the
form:
\begin{equation}
V^2 = \left[\frac{2J_1(\pi\Theta_{UD}(B/\lambda))} {\pi\Theta_{UD}(B/\lambda)
}\right]^2
\end{equation}
where $J_1$ is the first order Bessel function, $B$ is the projected baseline
length, $\lambda$ is the wavelength of the observations, and $\Theta_{UD}$ is
the apparent uniform disk angular diameter. The best fit uniform disk diameter
was found to be $ \Theta_{UD}= 0.527 \pm 0.028$ mas, ($\chi^2_\nu \sim 0.4$).

\begin{figure}[ht]

    \includegraphics[angle=90,scale=0.7,keepaspectratio=true]{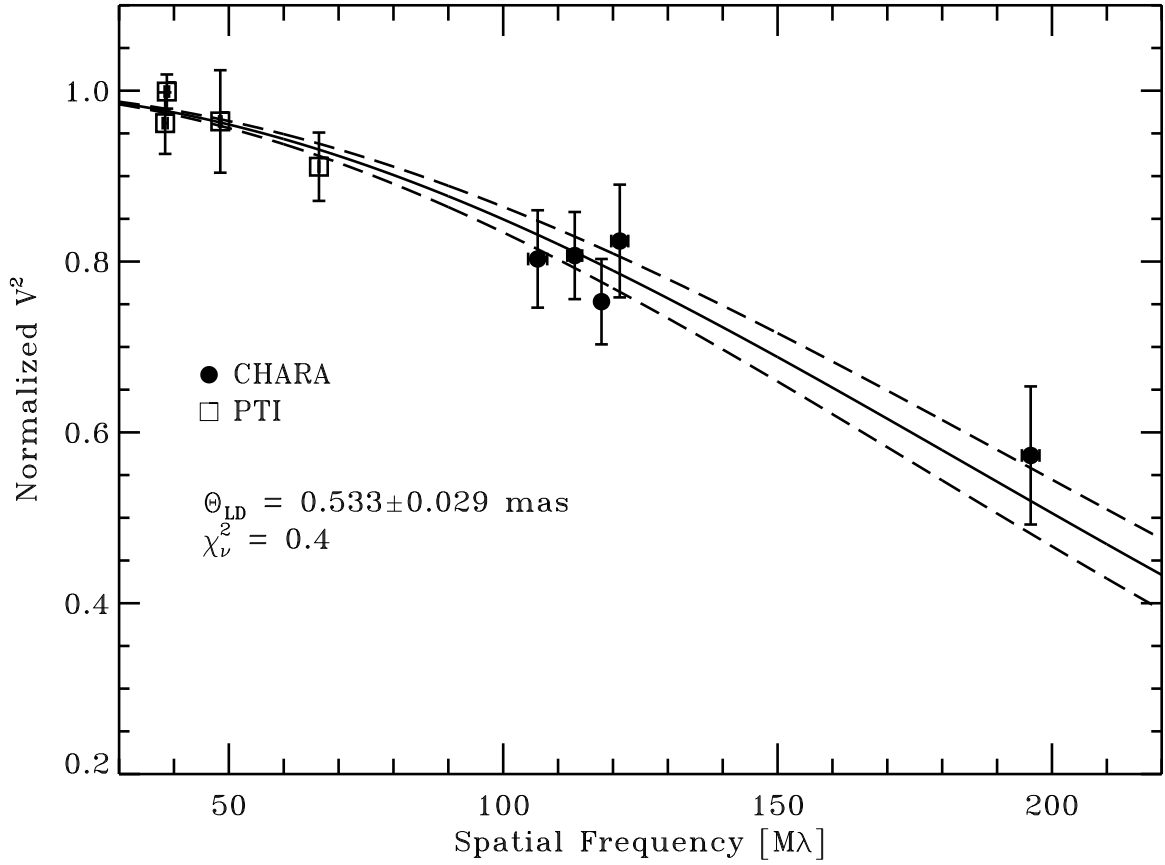}

    \figcaption{Normalized visibility vs. spatial frequency for $\lambda$ Boo
    as listed in Table \ref{meanv2-tab}.  Data obtained with the CHARA Array
    are shown with the filled circles; data obtained with PTI are shown with
    the open squares.  Error bars represent 1-$\sigma$ uncertainties. The
    solid line represents the best-fit limb-darkened stellar disk model fit.
    The dashed lines represent the 1-$\sigma$ fitting uncertainties.
    \label{v2-fig} }

\end{figure}

Limb darkening in A stars in the near-infrared is expected to be relatively
low \citep[e.g.][]{cdg95}; however, assuming that the star is a simple uniform
disk will cause an underestimation of the true, limb-darkened disk size of the
star by approximately 1\%. Assuming a linear limb darkening law, the
visibility function for a linear limb darkened stellar disk model can be
parameterized as:
\begin{equation}
V^2 = \left[\frac{1-\mu_{\lambda}}{2} + \frac{\mu_{\lambda}}{3}\right]^{-2}
\left[\frac{(1-\mu_{\lambda})J_1[\pi(B/\lambda)\Theta_{LD}]} {
\pi(B/\lambda)\Theta_{LD}} + \frac{(\mu_{\lambda})j_1[\pi(B/\lambda)
\Theta_{LD}]} {\pi(B/\lambda)\Theta_{LD}}\right]^2
\end{equation}
where $\mu_\lambda$ is the linear limb darkening coefficient($\mu \approx
0.16$ for $\lambda$~Boo; \citet{cdg95}), $j_1$ is the first order spherical
Bessel function, and $\Theta_{LD}$ is the apparent stellar limb darkened disk
angular diameter \citep{hb74}. The limb-darkening in the infrared for A-stars
is sufficiently small that a large change in $\mu_\lambda$ (25\%) results in a
very small change in the derived angular diameter ($\lesssim 0.5\%$). The best
fit limb darkened stellar disk diameter was determined to be $\Theta_{LD} =
0.533\pm0.029$ mas.  In Figure \ref{v2-fig}, we present the visibility curve
for $\lambda$~Boo with the best fit limb-darkened stellar disk model
overlayed, along with the 1-$\sigma$ model fitting boundaries.

The measured angular diameter is in agreement with the angular diameter as
predicted from interferometrically calibrated radius-color relationships for
single stars \citep[$\Theta_{predict}\approx 0.54-0.56$ mas;
][]{vanbelle99,kervella04}.  Speckle observations of $\lambda$~Boo
\citep{mca89} detected no companion brighter than $\Delta m \lesssim 2$ mag,
with a minimum separation of $0\farcs03$ ($30\ {\rm mas} \approx 1$ AU at the
distance of $\lambda$~Boo).  Further, Hipparcos observations of $\lambda$~Boo
display no signatures of a companion star or higher-order acceleration terms
in the parallactic solutions \citep{perryman97}. Finally, the interferometric
data presented here, spanning of nearly six years, are all consistent with a
single-star model (see Figure \ref{v2-fig}).

The interferometric data  do not represent a definitive null result for the
existence of a companion star to $\lambda$~Boo.  However, if $\lambda$~Boo
contains an unrecognized (i.e., unknowingly detected) binary companion
($\Delta\rm{K} \gtrsim 1.5-2$), the presence of a companion in the
interferometric data would {\em lower} the observed visibility amplitudes (as
compared to a single star) and lead to an {\em over-estimation} of the stellar
angular diameter. That, in turn, would imply that the true stellar radius is
{\em smaller} than observed. Thus, the single-star assumption leads to an
upper limit (within the measurement uncertainties) of the stellar radius.

\subsection{Radius and Mass \label{radmass-subsec}}

The parallax of $\lambda$~Boo, as measured by Hipparcos, is $\pi =
33.58\pm0.61$ mas \citep[$d = 29.78_{-0.53}^{+0.55}$
pc;][]{perryman97,hwp02}.  Taking the limb darkened stellar radius
as the Rosseland (photospheric) angular diameter, we derive a linear
radius for $\lambda$~Boo of $R_\star = 1.70\pm0.10\ R_\odot$.

If we combine the linear radius with a surface gravity, we can derive an
estimate for the mass of $\lambda$ Boo.   \citet{ck01} fit the IUE spectrum of
$\lambda$ Boo with an atmosphere model that is metal-poor in all the heavy
elements (${\rm[M/H]}=0.0$) except for C, N, \& O.  They found the best fit
model to have a temperature of 8500-8600 K and a surface gravity of $\log{(g)}
= 4.0$ $[\rm{cm\ s}^{-\rm{2}}]$.  They note that \citet{breger76}, by fitting
to only the visible part of the spectrum, determined a best fit temperature
and surface gravity of 8550 K and $\log{(g)} = 4.1$ $[\rm{cm\ s}^{-\rm{2}}]$.
Using photometric relationships, \citet{chen06} derive a surface gravity of
$\log{(g)} = 4.198$ $[\rm{cm\ s}^{-\rm{2}}]$.

From this surface gravity range, we infer a stellar mass range for
$\lambda$~Boo of $M_\star = 1.1-1.7 M_\odot$.  For a given surface gravity,
the linear radius uncertainty contributes approximately $\sigma(M_\star) =
0.1-0.2\ M_\odot$ to the total mass uncertainty. Thus, the uncertainty in the
mass (i.e., the range of masses derived) is primarily a result of the
uncertainty in the surface gravity.

In comparison, we have derived the masses for $\beta$ Leo (A3V), Sirius (A1V),
and Vega (A0V), three well-studied early A-type main sequence stars that have
had their diameters measured directly.  Of these three A-stars, $\beta$~Leo is
the closest to $\lambda$~Boo in spectral type (A3V {\it vs.} A3V kB9.5mB9.5),
and provides the best comparison to $\lambda$~Boo.

$\beta$ Leo and Sirius have limb darkened angular diameters of
$\Theta_{\beta\rm{Leo}} = 1.45\pm0.03$ mas \citep{difolco04} and
$\Theta_{\rm{Sirius}} = 6.01\pm0.02$ mas \citep{kervella03}. Combined with the
parallaxes $(\pi = 90.16\pm0.89\ \&\ 379.21\pm1.58$ mas), we derive linear
radii of $R_{\beta\rm{Leo}} = 1.72\pm0.04\ R_\odot$ and $R_{\rm{Sirius}} =
1.71\pm0.01\ R_\odot$ -- very similar to the radius measured for
$\lambda$~Boo.  With respective surface gravities of $\log{(g)} = 4.26$
$[\rm{cm\ s}^{-\rm{2}}]$ \citep{en03} and $\log{(g)} = 4.31$ $[\rm{cm\
s}^{-\rm{2}}]$ \citep{su89}, the derived masses of $\beta$ Leo and Sirius are
$M_{\beta\rm{Leo}} = 1.97\pm0.09\ M_\odot$ and $M_{\rm{Sirius}} = 2.01\pm0.05\
M_\odot$.  Vega is larger ($R \approx 2.5\ R_\odot$) and more massive
($M_{\rm{Vega}} = 2.3\pm0.2\ M_\odot$) than $\lambda$~Boo, $\beta$~Leo, and
Sirius \citep{aufdenberg06}.

\begin{figure}[ht]

    \includegraphics[angle=0,scale=0.9,keepaspectratio=true]{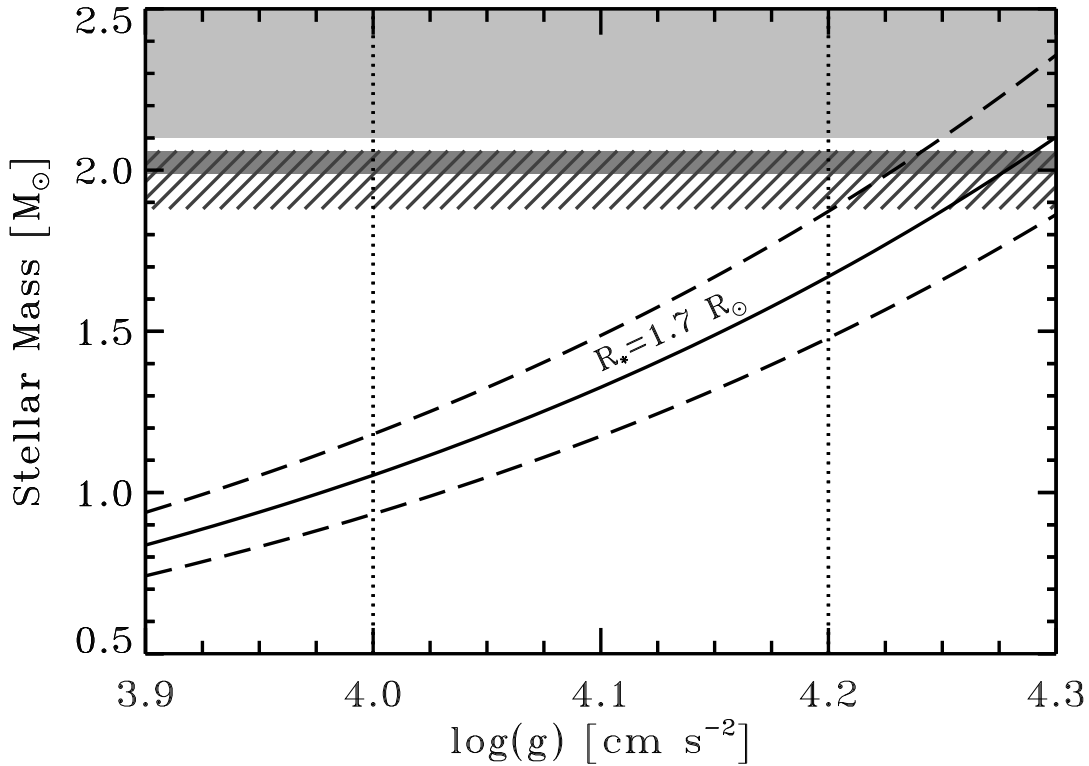}

    \figcaption{A plot of the derived stellar mass for $\lambda$~Boo as a
    function of surface gravity.  The solid line represents the linear radius
    ($R_\star = 1.7~R_\odot$) as derived from the measured angular diameter.
    The dashed lines represent the $1\sigma$ uncertainty limits for
    $\lambda$~Boo.  The vertical dotted lines delineate the range of surface
    gravity $(\log{(g)} = 4.0-4.2)$ for $\lambda$~Boo, as discussed in the
    text.  For comparison, the light gray region marks the derived mass of
    Vega; the dark gray region marks the derived mass for Sirius, and the
    hatched region marks the derived mass for $\beta$~Leo. \label{mass-fig} }

\end{figure}

The distribution of derived stellar mass as a function of surface gravity for
$\lambda$~Boo is shown in Figure \ref{mass-fig}. The figure demonstrates that
the mass for $\lambda$~Boo is in rough agreement (within $1\sigma$) with the
mass of $\beta$~Leo and Sirius if the surface gravity for $\lambda$~Boo is
$\log{(g)} \approx 4.2$. If the surface gravity is nearer to $\log{(g)} = 4.0$
or $\log{(g)} = 4.1$ as indicated by the detailed UV and optical spectral
fitting, then the derived mass for $\lambda$~Boo is $2-3\sigma$ below that
found for the three young A-stars $\beta$~Leo, Sirius, and Vega.

We note here that the known debris disk surrounding Vega was likely detected
with the interferometric observations at PTI \citep{ciardi01} and
independently with observations at the CHARA Array \citep{absil06}.
$\lambda$~Boo has a stronger mid-infrared excess than Vega, indicative of
circumstellar material surrounding the star which is the primary reason for
the conjectured association of $\lambda$~Boo stars with Vega-like stars
\citep[e.g.,][]{jura05,rieke05,chen06}. There is no evidence in our data that
the circumstellar material has been detected by the CHARA Array. However, {\em
if} the surrounding shell and/or disk indeed had been detected, the
circumstellar material would serve to make $\lambda$~Boo appear {\em larger}
than it actually is, yielding an upper limit to the stellar radius and mass.

\subsection{Evolutionary Status and Age \label{age-subsec}}

Previous estimates of the mass of $\lambda$~Boo have been made by placing it
on a luminosity-temperature HR diagram \citep{ib95, paunzen97, paunzen02} and
comparing its position to that of solar-metallicity stellar evolutionary
models \citep{schaller92, claret95, morel97}. These works report a $\lambda$
Boo effective temperature range of $T_{eff} \approx 8600 - 8900$ K and a
luminosity range of $L_{\star} \approx 15 - 24$ $L_\odot$ (see Table 4 in
\citet{paunzen02} for a summary).  The inferred mass range, from comparison to
the solar metallicity stellar evolutionary models, of these works is
$M_{\star} \approx 2.0 - 2.1$ $M_\odot$. We wish to place $\lambda$~Boo on a
luminosity-temperature HR diagram to explore the differences between our {\em
derived} mass for $\lambda$~Boo and the {\em inferred} mass by previous works.

The measured angular diameter allows us to derive the effective temperature of
$\lambda$~Boo via the Stefan-Boltzmann equation:
\begin{equation}
T_{eff} = \left[\frac{L_{\star}}{4\pi\sigma R_\star^2}\right]^{1/4} =
\left[\frac{F_{bol}D_\star^2}{\sigma R_\star^2}\right]^{1/4}
\end{equation}
where $L_\star$ is the luminosity, $R_\star$ is the stellar radius, $\sigma$
is the Stefan-Boltzmann constant, $F_{bol}$ is the bolometric flux and
$D_\star$ is the is the distance to the star. In terms of the angular diameter
in milli-arcsec ($\Theta$) and in units of $10^{-10}$~W~m$^{-2}$ for
$F_{bol}$, equation (3) may be written as
\begin{equation}
T_{eff} = 4163\left[\frac{F_{bol}}{\Theta^2}\right]^{1/4}.
\end{equation}

\begin{figure}[ht]

    \includegraphics[angle=90,scale=0.7,keepaspectratio=true]{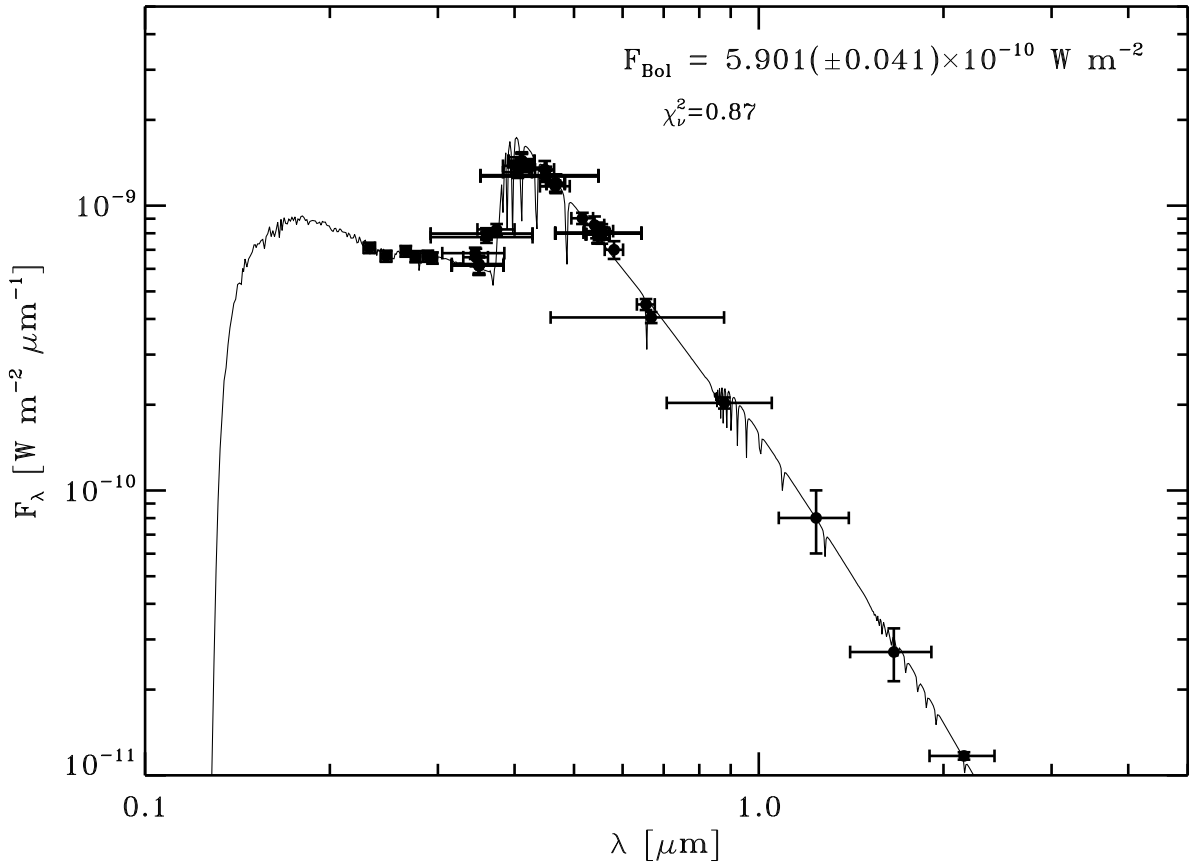}

    \figcaption{Model spectral energy distribution and flux density data for
    $\lambda$ Boo. The horizontal error-bars represent the bandwidths
    associated with the observations. The data have been fit with a 8750 K
    ${\rm[M/H]}=-2.0$ template from \citet{lcb97}. \label{sed-fig} }

\end{figure}

The bolometric flux for $\lambda$~Boo was estimated by fitting the
ultra-violet (IUE) to near-infrared (2MASS) spectral energy distribution with
templates from \citet{lcb97} (Figure \ref{sed-fig}).   The bolometric flux is
$F_{bol} = 5.901\pm0.041 \times 10^{-10}$~W~m$^{-2}$. At a distance of $d =
29.78_{-0.53}^{+0.55}$ pc, this corresponds to a luminosity of $L_\star = 16.3
\pm 0.6$~$L_\odot$. Combined with the limb darkened angular diameter, we
derive an effective temperature of $T_{eff} = 8887\pm242$~K. Our temperature
estimate is in good agreement with temperatures reported in the literature
which range from 8550~K \citep{ck01} to 8920~K \citep{hhk99}.

\begin{figure}[ht]
    \epsscale{0.7}

    \plotone{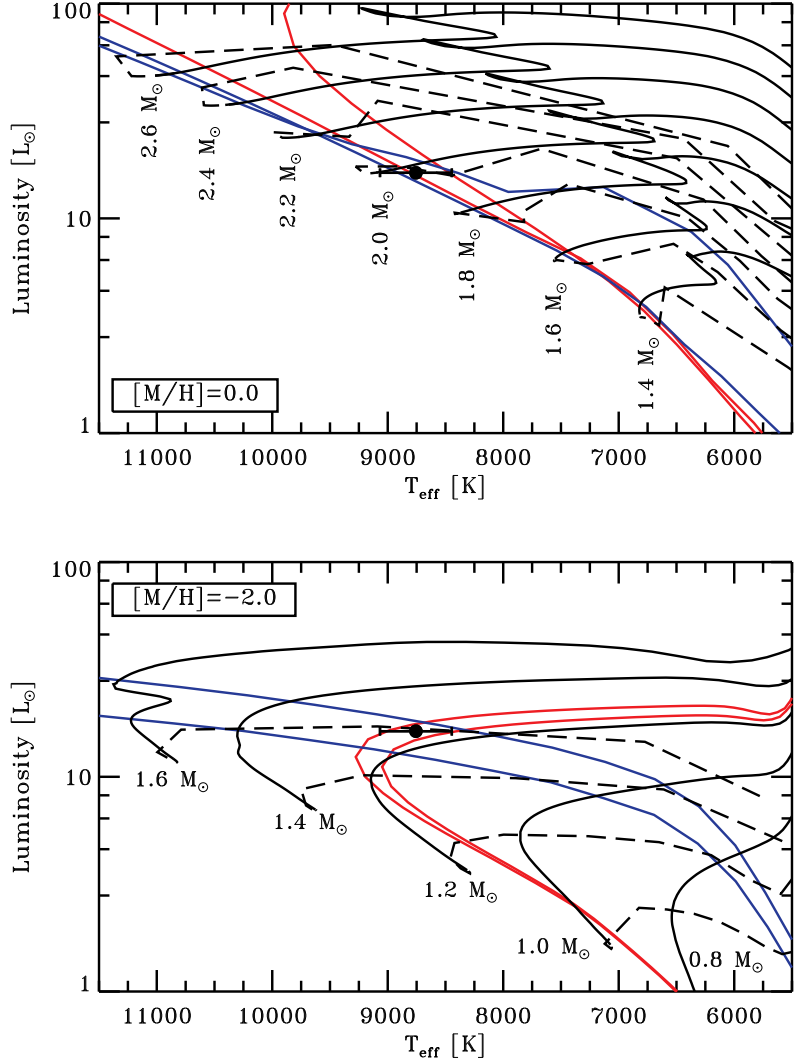}

    \figcaption{The derived linear radius and effective temperature for
    $\lambda$~Boo are shown versus the pre-main sequence evolutionary tracks
    (dashed black lines) and (post-)main sequence evolutionary tracks (solid
    black lines) for the Y$^2$ stellar evolutionary models. {\em Top}: Models
    for a metallicity of $[{\rm M}/{\rm H}] \approx 0.0$. {\em Bottom}: Models
    for a metallicity of $[{\rm M}/{\rm H}] \approx -2.0$. The stellar mass
    for each track is labelled in solar masses. The solid blue and solid red
    lines in each panel represent pre- and post-main sequence isochrones,
    respectively. For the solar metallicity models ({\em top}), the two
    pre-main sequence isochrones (blue) correspond to 8 Myr and 30 Myr, and
    the two post-main sequence isochrones (red) correspond to 80 Myr and 300
    Myr. For the metal-poor models ({\em bottom}), the pre-main sequence
    (blue) and post-main sequence (red) isochrones correspond to 3 \& 4 Myr
    and 2.6 \& 2.8 Gyr, respectively. \label{lumtemp-fig} }

\end{figure}

Using the interpolator provided with the the Yonsei-Yale (Y$^2$) stellar
evolutionary models \citep{demarque04}, we have generated isochrones and
evolutionary tracks for solar metallicity (z=0.02, ${\rm[M/H]}\approx 0.0$)
and sub-solar metallicity (z=0.0002, ${\rm[M/H]}\approx -2.0$).  For the mass
tracks, the stellar masses span $0.8 - 2.7\ M_\odot$ in steps of $0.1\
M_\odot$, evolved across both the pre-main sequence and post-main sequence.
The HR diagrams, in terms of stellar luminosity vs. effective temperature as
represented by the Y$^2$ models, are shown in Figure \ref{lumtemp-fig}.  The
position of $\lambda$~Boo, as measured by the interferometers, is marked in
each of the HR diagrams.

The position of $\lambda$~Boo on the solar-metallicity diagram ({\em top}
Figure \ref{lumtemp-fig}) implies that $\lambda$~Boo should have a stellar
mass of $M_\star = 1.9-2.1\ M_\odot$, in agreement with the upper bound
derived for the mass of $\lambda$~Boo ($\log(g)=4.2,\ M_\star = 1.7\pm0.2\
M_\odot$). If $\lambda$~Boo is represented by the solar metallicity models,
the star is fairly young with an age of $8 - 300$ Myr. This age would be
consistent with $\lambda$~Boo being related to the Vega-like stars (i.e.,
stars with dusty debris disks), but being younger than Vega itself
\citep{jura05,rieke05}.

In contrast, placing $\lambda$~Boo on a set of sub-solar metallicity models
({\em bottom} Figure \ref{lumtemp-fig}), the (post)-main sequence models imply
a stellar mass of $M_\star = 1.2-1.4\ M_\odot$, in agreement with the
mid-range for the mass derived from our observations ($\log(g)=4.1,\ M_\star =
1.3\pm0.2\ M_\odot$). The pre-main sequence tracks imply a slightly larger
stellar mass of $M_\star = 1.5-1.6\ M_\odot$.

The ages associated with the sub-solar metallicity pre- and post-main sequence
tracks are quite different from each other. For these models, if $\lambda$~Boo
is a pre-main sequence star, it would need to be extremely young ($3-4$ Myr).
At such a young age, the star should be associated with the Herbig AeBe
(HAeBe) stars. Yet, $\lambda$~Boo shows no Balmer emission lines
\citep[e.g.,][]{ib98}, an observational requirement of the HAeBe stars
\citep{the94}. Additionally, at a galactic position of $l=86^\circ,
b=65^\circ, d=29.8\ {\rm pc}$, $\lambda$~Boo is not directly associated with
any molecular clouds or regions of high extinction \citep{lucke78,gvb93}.
These discrepancies suggest that $\lambda$~Boo, if best described by the
sub-solar metallicity evolutionary tracks, is not $3-4$ Myr old.

If $\lambda$~Boo is a post-main sequence star, the sub-solar metallicity
models place it at an age of $2-3$ Gyr, and the star is at (or past) the
terminal age for the main sequence. We note here that this independent
assessment of the age of $\lambda$~Boo is in general agreement with the
results of \citet{paunzen02} who found that the $\lambda$~Boo stars span an
age range of 10 Myr to 1.5 Gyr with a strong peak near 1.0 Gyr.

\section{Summary \label{sum-sec}}

We have presented the first direct determination of the angular size of the
chemically peculiar star $\lambda$ Bo\"otis.  The infrared interferometric
observations made use of the longest baselines on the CHARA Array and the
Palomar Testbed Interferometer. The primary result of this work is the direct
determination of the limb darkened angular diameter of $\lambda$~Boo, which
was measured to be $\Theta_{LD} = 0.533\pm0.029$ mas. A full summary of the
stellar parameters derived from the spatially resolved interferometric
observations are presented in Table \ref{starparam-tab}.

In combining our independently determined stellar radius with previous
determinations of the surface gravity, we have calculated a stellar mass range
for $\lambda$~Boo of $M_\star = 1.1-1.7\ M_\odot$.  The radius determination
contributes $0.1-0.2\ M_\odot$ to the uncertainty.  The remainder of the mass
uncertainty is contributed entirely by the uncertainty in surface gravity
($\log{(g)} = 4.0-4.2$).

Solar-metallicity (z=0.02, ${\rm[M/H]}\approx 0.0$) stellar evolutionary
models predict that $\lambda$~Boo should have a mass nearer to $1.9-2.1\
M_\odot$, in agreement with the upper bound of our mass determination
($\log(g)=4.2,\ M_\star = 1.7\pm0.2\ M_\odot$). Metal-poor (z=0.0002,
${\rm[M/H]}\approx -2.0$) stellar evolutionary models predict a mass $1.2-1.4\
M_\odot$ in agreement with the mid-range of our interferometrically derived
mass ($\log(g)=4.1,\ M_\star = 1.3\pm0.2\ M_\odot$).  A more definitive
surface gravity determination is required to distinguish between these two
sets of models.

\acknowledgments

The authors would like to thank the entire staff at the CHARA Array for
without them this work would not have been possible. Portions of this work
were performed at the California Institute of Technology under contract with
the National Aeronautics and Space Administration. This research has been
supported by National Science Foundation grants AST-0307562 and AST-0606958 to
Georgia State University. Additional support has been received from the
Research Program Enhancement program administered by the Vice President for
Research at Georgia State University. Work done with the Palomar Testbed
Interferometer was performed at the Michelson Science Center, California
Institute of Technology, under contract with the National Aeronautics and
Space Administration. Interferometer data were obtained at Palomar Observatory
using the NASA PTI, supported by NASA contracts to the Jet Propulsion
Laboratory. Science operations with PTI are conducted through the efforts of
the PTI Collaboration, and we acknowledge the invaluable contributions of our
PTI colleagues.

This research has made use of the NASA/IPAC Infrared Science Archive, which
is operated by the Jet Propulsion Laboratory, California Institute of
Technology, under contract with the National Aeronautics and Space
Administration. This research has made use of data obtained from the High
Energy Astrophysics Science Archive Research Center (HEASARC), provided by
NASA's Goddard Space Flight Center.  This research has made use of the SIMBAD
database, operated at CDS, Strasbourg, France.

\newpage

\begin{deluxetable}{cccc}
\tablecolumns{4}
\tablewidth{0pc}
\tablecaption{Calibration Stars \label{calib-tab} }
\tablehead{ \colhead{Star} & \colhead{$\theta_{EST}$\tablenotemark{a}} &
\colhead{Distance from} & \colhead{Spectral}\\
\colhead{} & \colhead{(mas)} & \colhead{$\lambda$~Boo (deg)} & \colhead{Type}
} \startdata
HD 129002 & $0.198 \pm 0.012$ & 4.3 & A1 V \\
HD 125349 & $0.286 \pm 0.018$ & 5.3 & A1 IV \\
\enddata
\tablenotetext{a}{Estimated angular diameters derived from spectral
energy distribution modeling.}
\end{deluxetable}


\begin{deluxetable}{ccrccc}
\tablecolumns{5} \tablecaption{Weighted Mean Visibilities
\label{meanv2-tab}}
\tablewidth{0pt}
\tablehead{ & \colhead{Number of} & \colhead{Mean\tablenotemark{a}} &
\colhead{Mean\tablenotemark{a}}
& \colhead{Mean\tablenotemark{a}} & \colhead{Mean\tablenotemark{b}}\\
\colhead{Array} & \colhead{Points in} & \colhead{Projected} &
\colhead{Position Angle} & \colhead{Effective} & \colhead{Normalized}\\
& \colhead{Average} & \colhead{Baseline} & \colhead{E. of N.} &
\colhead{Wavelength} & \colhead{V$^2$}
\\
&  & \colhead{$(\rm{m})$} & \colhead{$(\rm{deg})$} & \colhead{($\mu$m)}
 }
\startdata
CHARA & 3 & 226.7 (3.7) & 291.0 (1.1) & 2.133 & $0.803\pm0.057$\\
CHARA & 4 & 241.0 (1.1) & 295.6 (1.0) & 2.133 & $0.897\pm0.051$\\
CHARA & 3 & 251.5 (1.8) & 299.8 (0.8) & 2.133 & $0.753\pm0.050$\\
CHARA & 3 & 258.6 (3.3) & 303.3 (2.6) & 2.133 & $0.824\pm0.066$\\
CHARA & 3 & 328.1 (2.0) & 191.8 (4.0) & 1.673 & $0.573\pm0.081$\\
  PTI & 20 &  85.8 (0.9) & 241.9 (7.7) & 2.217 (0.005) & $0.999\pm0.020$\\
  PTI & 11 &  85.0 (1.0) & 325.7 (7.6) & 2.214 (0.002) & $0.962\pm0.036$\\
  PTI & 25 & 108.5 (0.6) & 191.3 (6.6) & 2.242 (0.008) & $0.965\pm0.040$\\
  PTI & 6 & 109.0 (0.1) & 179.6 (6.1) & 1.641 (0.001) & $0.911\pm0.040$\\
\enddata
\tablenotetext{a}{Values in parentheses represent the rms
dispersions.}
\tablenotetext{b}{Uncertainties derived from the
weighted mean.}
\end{deluxetable}

\begin{deluxetable}{ccccc}
\tablecolumns{3}
\tablewidth{0pc}
\tablecaption{$\lambda$ Bo\"otis Stellar Properties \label{starparam-tab} }
\tablehead{
\colhead{Parameter} & \multicolumn{2}{c}{Value Using}  & Units & Reference\\
& \colhead{$[{\rm M}/{\rm H}]=0.0$ } & \colhead{$[{\rm M}/{\rm
H}]=-2.0$} }
\startdata
Parallax  & \multicolumn{2}{c}{$33.58\pm0.61$} & mas & 5 \\
Limb Darkened Diameter & \multicolumn{2}{c}{$0.533\pm0.029$} & mas & 1 \\
Linear Radius & \multicolumn{2}{c}{$1.70\pm0.10$} & R$_\odot$ & 1 \\
$v \sin{(i)}$ & \multicolumn{2}{c}{$100\pm10$} & km s$^{-1}$ & 4\\
Bolometric Flux & \multicolumn{2}{c}{$5.901\pm0.041$} & $10^{-10}$ W
m$^{-2}$ & 1 \\
Luminosity & \multicolumn{2}{c}{$16.3\pm0.6 $} & L$_\odot$ & 1\\
Effective Temperature &\multicolumn{2}{c}{$8887\pm242$}& K & 1 \\
Surface Gravity & \multicolumn{2}{c}{$4.0 - 4.2$} & $\log{[{\rm cm\ s}^{-1}}]$ & 2,3 \\
Mass & \multicolumn{2}{c}{$1.1 - 1.7$\tablenotemark{a}} & M$_\odot$ & 1 \\
Pre-MS Age & $8-30$ & $3-4$ & Myr & 1\\
Post-MS Age & $0.08-0.3$ & $2-3$ & Gyr & 1\\
Model Mass Range & $1.9-2.0$ & $1.3-1.6$ & M$_\odot$ & 1
\enddata
\tablenotetext{a}{The radius uncertainty contributes approximately an
uncertainty of 0.1-0.2 $M_\odot$ for a given value of the surface gravity.
The range in mass represents the range in surface gravity.} \tablerefs{ 1.
This Work; 2. \citet{ck01}; 3. \citet{chen06}; 4. \citet{hwp02}; 5.
\citet{perryman97} }

\end{deluxetable}

\clearpage

\end{document}